\documentclass[prb,reprint,twocolumn,preprintnumbers,amsmath,amssymb,showpacs,footinbib,superscriptaddress]{revtex4-1}

\usepackage[titletoc,toc,title]{appendix}
\usepackage{braket}
\usepackage[final]{feynmp}
\usepackage{ifpdf}
\usepackage{comment}
\usepackage{amsmath}
\usepackage{mathrsfs}
\usepackage{color}
\usepackage{bm}
\usepackage[font=small]{subcaption}
\usepackage[font=small]{caption}

\graphicspath{{./Figures/}}

\makeatletter
\def\endfmffile{%
	\fmfcmd{\p@rcent\space the end.^^J%
			end.^^J%
			endinput;}%
	\if@fmfio
		\immediate\closeout\@outfmf
	\fi
	\ifnum\pdfshellescape=\@ne
		\immediate\write18{mpost \thefmffile}%
	\fi}
\makeatother

\usepackage[demo]{graphicx}
\usepackage{dcolumn}
\usepackage{bm}
\usepackage{hyperref}

\newcommand{\beq}{\begin{equation}}
\newcommand{\eeq}{\end{equation}}

\renewcommand{\thanks}[1]{\footnote{#1}} 

\newcommand{\be}{\begin{equation}}
\newcommand{\bea}{\begin{eqnarray}}
\newcommand{\eea}{\end{eqnarray}}

\newcommand{\ee}{\end{equation}}

\def\ba{\begin{eqnarray}}
\def\ea{\end{eqnarray}}

\def\14{{1\over4}}
\def\12{{1 \over 2}}

\def\8{\infty}

\def\d{\partial}

\def\undertext#1{\vtop{\hbox{#1}\kern 1pt \hrule}}

\def\VEV#1{\left\langle\,#1\,\right\rangle}

\def\be{\begin{equation}}
\def\ee{\end{equation}}
\def\bea{\begin{eqnarray} & &}
\def\eea{\end{eqnarray}}

\def\rf#1{(\ref{#1})}

\def\rf#1{(\ref{#1})}

\begin{document}


\title{Randomly measured quantum particles and thermal noise}

\author{Victor Gurarie}
\affiliation{Department of Physics and Center for Theory of Quantum Matter, University of Colorado, Boulder, Colorado 80309, USA}

\begin{abstract}
We consider the motion of a quantum particle whose position is measured in random places at random moments in time. 
We contrast this motion with the motion of a quantum particle in a potential which varies randomly in space and in time, which could also be thought of as  (possibly thermal) noise. 
We calculate expectations of observables both linear and nonlinear in the density matrix. We demonstrate explicitly that while linear observables cannot distinguish between random measurements and random noise,
measurable distinctions can be seen in nonlinear observables. 
\end{abstract}

\maketitle



It became widely appreciated in recent years that dynamic evolution of a quantum system, when it is subject to  measurements, acquires new features \cite{Fisher2018,Nahum2019}.
 Unlike the Schr\"odinger evolution of a state of a quantum particle, described by a unitary evolution operator, measurements evolve the wave function according to Born's rules, that is by application of nonunitary projection operators to the wave function. Evolution of the wave function by these abrupt projections are difficult to capture analytically. Instead, weak measurement operators can be introduced, describing the 
 process where a system is coupled to an ancilla which is then projectively measured. 
 The only condition that the weak measurement operators must satisfy is the Kraus operator condition
 \be \label{eq:kraus} \sum_j \hat K_j^\dagger \hat K_j = 1,
 \ee
 where the index $j$ labels the outcomes of the measurement. 
 These operators can be chosen to be close to identity, thus providing us with wave function evolution continuous in time \cite{Murciano2023,Garratt2023,Poboiko2023,Guo2024}.  
 
 For example, for a particle whose position is measured at random moments in time can be described by the Kraus operator 
 \be \label{eq:kv} \hat K \left[ V \right] = \int dx \, e^{V(x) \delta t} \left| x\right> \left< x \right|,
 \ee
 where $V(x)$ is a function corresponding to the outcomes of the measurement procedure (thus it replaces the index $j$ above) and $\delta t$ is the time interval between the measurements. 
 In particular, choosing $V(x) \delta t = 1$ for $a<x<b$ and $V(x) \rightarrow -\infty$ otherwise reduces $\hat K[V]$ to a projection operator onto the interval $[a,b]$, so its application to the wave function describes the
 measurement outcome where the particle is confined to this interval. 
 
If instead $V(x)$ remains arbitrary, 
 the limit of small $\delta t$ turns $\hat K[V]$ into an operator very close to identity and allows for continuous evolution in time.  Summation over $j$ is now replaced by the integration over the functions $V$.

Taking the limit of continuous time evolution results in the Schr\"odinger-like
 evolution equation
 \be \label{eq:s1} i \frac{\partial \psi}{\partial t} = -\frac{1}{2m} \frac{\partial^2 \psi}{\partial x^2} + i V(x,t) \psi,
 \ee
 with $V(x,t)$ random in space and time, which can be taken to be a Gaussian random function with the average
 \be \label{eq:av}  \VEV{V(x_1,t_1) V(x_2, t_2) } = \lambda \, \delta(t_1-t_2) \, W(x_1-x_2).
 \ee 
 Importantly, when summation over the discrete index $j$ is replaced by averaging over Gaussian random functions $V$, the operator $\hat K[V]$ defined in \rf{eq:kv} satisfies the
 Kraus condition \rf{eq:kraus}, up to an overall constant which can be removed with proper normalization.
 We can take $W$ in the form
 \be \label{eq:W} W(x) = \frac 1 \ell \,  g \left( \frac x \ell \right),
 \ee where 
 \be \int d\xi \, g(\xi) =1.
 \ee
 $\ell$ can be thought of as a length of the interval where the particle is measured. $\ell \rightarrow 0$ limit corresponds to $g$ reducing to a delta-function. 
 
 A randomly measured quantum particle can be contrasted with a quantum particle moving in a potential, random in both space and time (a particle subject to noise, possibly thermal \cite{kubo1966fluctuation,landau1980statistical,Kamenev2011}), described by the
 time-dependent Schr\"odinger equation
  \be \label{eq:s2} i \frac{\partial \psi}{\partial t} = -\frac{1}{2m} \frac{\partial^2 \psi}{\partial x^2} +  V(x,t) \psi,
 \ee
 The only difference between the equations \rf{eq:s1} and \rf{eq:s2} is the imaginary unit in front of the random potential making the evolution described by the equation \rf{eq:s1} nonunitary. 
 Both setups involve a random potential, so to distinguish them we will call \rf{eq:s1} random measurements and \rf{eq:s2} random noise. 
 
 In what follows we show that the expectation values of operators when averaged over random measurements such as encoded in the Schr\"odinger equation \rf{eq:s1} coincide with the expectation values of operators averaged over noise, as encoded in \rf{eq:s2}. In particular, quantum mechanical average of the square of the position of the particle, further averaged over either random measurements or noise, as a function of time $t$ is given by 
 \rf{eq:res}. To be able to distinguish between random measurements and random noise, we need look at higher powers of the quantum mechanical averages of operators, subsequently averaged over either measurements or noise. A quantum mechanical average of the position operator, which is then squared and averaged over measurements or noise, is given by either \rf{eq:meas} or \rf{eq:noise} respectively.

 
 Here is how we derive these results. If we are interested in measuring expectation value of some operator, such as the position operator of the particle, at a certain time $t$, we construct the density matrix $\rho(x_+,x_-,t)=\psi(x_+,t) \psi^*(x_-,t)$ averaged over the
 random potential $V$. The averaging over the random potential can be done explicitly, which is especially easy to carry out in Feynman's path integral approach to quantum mechanics. Indeed, 
 the wave function is given by (in case of random measurements)
 \be \label{eq:mf}  \psi(y_+,t) = \int_{x_+(t)=y_+} {\cal D} x_+(\tau) \, e^{i \int d\tau \left( \frac{m \dot x_+^2}{2} - i V(x_+) \right)},
 \ee
 while its complex conjugate is given by the complex conjugate counterpart of this functional integral
 \be \label{eq:mb} \psi^*(y_-,t) = \int_{x_-(t)=y_+} {\cal D} x_-(\tau) \, e^{-i \int d\tau \left( \frac{m \dot x_-^2}{2} + i V(x_-) \right)}.
 \ee
 The $x_+(\tau)$ and $x_-(\tau)$ trajectories are often referred to  as belonging to the forward and backward Keldysh contours \cite{Kamenev2011}. 
 
 To average the density matrix, we multiply these together and use
 \be \VEV{e^{\int d\tau V(x_+) + \int d\tau V(x_-)}} = e^{\lambda \int d\tau \left(W(x_+-x_-) + W(0)\right)},
 \ee
 where brackets denote averaging over the random potential $V$ by using \rf{eq:av}. Note that $W(0)$ is just an irrelevant constant which can be dropped. This immediately leads to the  equation satisfied by the density matrix (it could be called a Schr\"odinger-like equation, or a Schr\"odinger equation for the density matrix in the Choi-Jamio\l kowski  representation \cite{Jiang2013}, or a Liouville equation)
 \be \label{eq:schr2}  i \dot \rho =  - \frac{1}{2m} \frac{\partial^2 \rho}{\partial x_+^2} + \frac{1}{2m} \frac{\partial^2 \rho}{\partial x_-^2}  + i \lambda W(x_+-x_-) \rho.
\ee
 If alternatively we were interested in averaging over noise, as opposed to random measurements, we would write down
 \be \label{eq:nf} \psi(y_+,t) = \int_{x_+(t)=y_+} {\cal D} x_+(\tau) \, e^{i \int d\tau \left( \frac{m \dot x_+^2}{2} -  V(x_+) \right)},
 \ee
 \be \label{eq:nb} \psi^*(y_-,t) = \int_{x_-(t)=y_+} {\cal D} x_-(\tau) \, e^{-i \int d\tau \left( \frac{m \dot x_-^2}{2} - V(x_-) \right)}.
 \ee
 The averaging now leads to
\be \VEV{e^{i \left( \int d\tau V(x_-) - \int d\tau V(x_+) \right)}} = e^{\lambda \int d\tau \left(W(x_+-x_-) - W(0)\right)},
 \ee
which in turn implies that the density matrix satisfies exactly the same equation \rf{eq:schr2}, since $W(0)$, as was already pointed out, is an irrelevant constant. 
 
We now come to a striking conclusion that observables averaged over measurements or averaged over a random potential  are indistinguishable from each other. 

Now the equation \rf{eq:schr2} can be solved quite generally. Suppose initially at $t=0$ the particle is confined to  a particular region in space, $\psi_0 (x) =\exp\left(-x^2/(4 \Delta^2) \right)/\left(2 \pi \Delta^2 \right)^{1/4}$. With this initial wave function, 

\be \left< \psi_0 \right| \hat x^2 \left| \psi_0 \right>=\int dx \, \rho(x,x,0) \, x^2 =\Delta^2. 
\ee
In other words, we need to solve equation \rf{eq:schr2} with the initial conditions
\be \label{eq:init} \rho(x_+,x_-,0) = \frac{1}{\sqrt{2 \pi} \Delta}  \exp\left( - \frac {x_+^2+x_-^2}{4 \Delta^2} \right).
\ee
The solution can be found analytically for any function $W$. In particular it can be shown that at a time $t$ the average square of the position of the particle is given\cite{Rosenbluth1992,Gurarie2025} by (see also Appendix \ref{sec:a})
\be \label{eq:res} \int dx \, \rho(x,x,t) \, x^2 = \Delta^2 + \frac{\hbar^2 }{4 m^2 \Delta^2} t^2 + \frac{ \lambda \left| W''(0) \right| }{3 m^2} t^3.
\ee
Here to further elucidate the meaning of this answer  the Planck constant $\hbar$ was explicitly restored. 

The term in this expression proportional to $t^2$ is the standard quantum spreading of the wave packet. That spread is ballistic with $x^2 \sim t^2$, and is proportional to $\hbar^2$, consistent with its quantum nature. 
The second term where $x^2 \sim t^3$ is the result of the random measurements or random potential. Interestingly enough, that term has no Planck constant $\hbar$ so it is essentially classical. 
And indeed, this expression can be derived classically. The correspond to a particle which receives kicks, either because it is measured and confined to an interval of length $\ell$ making its velocity change abruptly due to uncertainty principle, or because it receives kicks directly from a random force due to the random noise. 

The result $x^2 \sim t^3$ in the context of random noise has already been obtained some time ago \cite{Golubovic1991,Rosenbluth1992}. 

We would now like to examine if there are any measurable differences between  random measurements and random noise. Those become explicit if we look at the correlators which are higher order in density matrix. While highly nonlinear correlators such as entanglement entropy has been popular to study in the context of random measurements, recently some simpler although still nonlinear observables were introduced
\cite{Vasseur2022,Huse2022}. For example, for a given function $V(x,t)$ one can compute the quantum mechanical average position $\VEV{x}$ and then average the square of this over the random potential (or the outcomes of measurements)
\cite{Gefen2023}. In what follows, we will call this quantity $\overline{\VEV{x}^2}$. 

To accomplish this we need to replicate the system by introducing $n$ copies of it. As has been pointed out in many publications, in the context of measurements we then need to take the limit $n \rightarrow 1$. Let us
see how this limit comes about. After the measurement operator $\hat K_\alpha$ is applied to the wave function, it needs to be normalized. A square of the quantum mechanical average of the operator $\hat x$, in turn
averaged over the outcome of measurements, can be written as
\be \label{eq:x2} \overline{\VEV{x}^2}=  \sum_j \left< \psi \right| \hat K_j^2 \left| \psi \right> \left(  \frac{\left<  \psi \right| \hat K_j  \hat x \hat K_j \left| \psi \right>}{\left< \psi \right| \hat K_j^2 \left| \psi \right>} \right)^2.
\ee
Here the factor $\left< \psi \right| \hat K_j^2 \left| \psi \right>$ plays a dual role: in the denominator it normalizes the wave function after application of the operator $\hat K_j$ while this same 
factor in front of the expression it gives the Born probability of the outcome of the measurement, so that summation over $j$ would represent averaging over measurement outcomes. 
We can obtain this expression also by replicating the system $n$ times and writing 
\begin{eqnarray} \label{eq:x22} \overline{\VEV{x}^2} &=&  \lim_{n \rightarrow 1} \sum_j \left<  \psi_1 \right| \hat K_j  \hat x \hat K_j \left| \psi_1 \right>\left<  \psi_2 \right| \hat K_j  \hat x \hat K_j \left| \psi_2 \right> \times \cr &&
\prod_{\alpha=3}^n \left<  \psi_\alpha \right| \hat K_j^2    \left| \psi_\alpha \right>.
\end{eqnarray}
It is easy to see that \rf{eq:x22} is equivalent to \rf{eq:x2} since
\begin{eqnarray} && \lim_{n \rightarrow 1} \prod_{\alpha=3}^n \left<  \psi_\alpha \right| \hat K_j^2    \left| \psi_\alpha \right> = \lim_{n \rightarrow 1} \left( \left<  \psi \right| \hat K_j^2    \left| \psi \right> \right)^{n-2} =
\cr &&
\frac{1}{\left<  \psi \right| \hat K_j^2    \left| \psi \right>}.
\end{eqnarray}
Note that in the context of random potential (noise), the wave function is already automatically normalized so that the limit $n \rightarrow 1$ is actually not needed. 

To implement this in the context of a randomly measured particle, we set up a replicated system with $2n$ coordinates $x_\alpha^+$ and $x_\alpha^-$, $\alpha = 1, 2, \dots, n$.  Multiplying $n$ functions $\psi(x_\alpha^+)$ constructed according to \rf{eq:mf} and $n$ functions $\psi^*(x_{\alpha}^-)$ constructed according to \rf{eq:mb} together (here $\alpha = 1, 2, \dots, n$) and averaging over random $V$ produces the following effective evolution equation for the replicated density matrix
\ \begin{eqnarray} \label{eq:schr3} && i \dot \rho + \frac{1}{2m} \sum_{\alpha=1}^n \left( \frac{\d^2 \rho}{\d {x_\alpha^+}^2} - \frac{\d^2 \rho}{\d {x_\alpha^-}^2} \right) = \cr &&  \frac{i \lambda}{2} \sum_{\alpha, \beta=1 }^n  2 W\left(x_{\alpha}^+ - x_{\beta}^- \right) \rho  \, +  \cr &&  \frac{ i \lambda } 2 \sum_{\alpha, \beta=1}^n \left[ W\left(x_{\alpha}^+ - x_{\beta}^+ \right) +W\left(x_{\alpha}^- - x_{\beta}^- \right)  \right] \rho.  \end{eqnarray}
 Alternatively if we are interested in studying noise (random potential), we will multiply together functions constructed according to \rf{eq:nf} and \rf{eq:nb}, to find the equation describing motion with noise
 \begin{eqnarray} \label{eq:schr4} && i \dot \rho + \frac{1}{2m} \sum_{\alpha=1}^n \left( \frac{\d^2 \rho}{\d {x_\alpha^+}^2} - \frac{\d^2 \rho}{\d {x_\alpha^-}^2} \right) = \cr && \frac{ i \lambda }2 \sum_{\alpha, \beta=1 }^n  2 W\left(x_{\alpha}^+ - x_{\beta}^- \right) \rho \,  -  \cr && \frac{  i \lambda }2  \sum_{\alpha, \beta=1}^n \left[ W\left(x_{\alpha}^+ - x_{\beta}^+ \right) +W\left(x_{\alpha}^- - x_{\beta}^- \right)  \right] \rho.  \end{eqnarray}
The only difference between equations \rf{eq:schr3} and \rf{eq:schr4} is in the sign in front of the terms which couple different replicas in the same Keldysh sector. While that difference did not matter when $n=1$, as was emphasized earlier, once we have more than one replica measurements and noise become qualitatively distinct. 

We would now like to solve the equations \rf{eq:schr3} and \rf{eq:schr4} to construct, for example, the square of the quantum mechanical average of the particle's position, subsequently averaged over measurement outcomes. To do that, we need to find $\rho$ and calculate
\be \label{eq:res2} \overline{\VEV{x}^2} = \lim_{n \rightarrow 1} \int \left[ \prod_{\alpha=1}^n d x_\alpha \right] x_1 x_2 \, \rho(x_1, x_1, x_2, x_2, \dots, x_n, x_n).
\ee
In the integral above, $x_\alpha^+ = x_\alpha^- = x_\alpha$ for all $\alpha$. 

Unfortunately these Schr\"odinger-like equations do not appear to be integrable and cannot be solved exactly for arbitrary $n$, even if the functions $W$ are replaced by the delta-functions. Nevertheless we can solve them perturbatively in powers of $\lambda$ (which has the meaning of the rate of measurements, or the strength of the random noise). This is motivated by the answer for the average square of the position of the particle \rf{eq:res}, which is linear in $\lambda$. Therefore, for this quantity perturbation theory in $\lambda$ is actually exact. While we have no reasons  to believe that for the quantity  \rf{eq:res2} the perturbation theory would be exact, in the absence of any other  available technique we proceed with perturbation theory. We might not be able to fully access possible regimes of randomly measured particle, or particle moving in a noisy environment, but we will be able to see the qualitative difference between the two at least when $\lambda$ is small. 

The actual calculation is straightforward. We write $\rho=\rho_0+\rho_1$, where $\rho_0$ solves the equations \rf{eq:schr3} or \rf{eq:schr4} without the right hand side with the initial condition \rf{eq:init}, while $\rho_1$ 
solves those same equations with $\rho\rightarrow \rho_1$ on the left hand side and $\rho \rightarrow \rho_0$ on the right hand side. By construction, $\rho_1$ is proportional to $\lambda$. We then use $\rho_1$ to evaluate \rf{eq:res2} (the contribution of $\rho_0$ to it vanishes). 

To simplify the algebra, we choose a convenient Gaussian form for the function $W$, by choosing the function $g$ in \rf{eq:W} as
\be \label{eq:g} g(\xi) = \frac{1}{\sqrt{2\pi}} e^{- \frac{\xi^2}{2}}.
\ee
The calculations are straightforward but tedious and so are described in  the Appendix \ref{sec:a}. The end result for the solution of the random measurement equation \rf{eq:schr3} is
\be \overline{\VEV{x}^2} = \frac{ \lambda t^2}{2 \sqrt{\pi } \Delta  m} \int_0^1 dx \frac{ \left(x -\frac{4 \Delta ^4 m^2}{t^2} \right)^2}{ \left(\frac{2 \Delta^2
   m^2 \left(2 \Delta ^2+\ell ^2\right)}{t^2} +x^2\right)^{3/2}}.
\ee
While this integral can be calculated analytically, the resulting expression is not very illuminating. Let us just look at it for small time $t \ll m \Delta^2$, where it reduces to
\be \label{eq:meas} \overline{\VEV{x}^2}    \approx \frac{2 \sqrt{2} \Delta^4 \lambda t}{\hbar^2 \sqrt{\pi} \left( 2 \Delta^2+ \ell^2 \right)^{3/2}}. \ee 
Here again for clarity we restored the Planck constant $\hbar$. Unlike the answer for the quantity $\VEV{x^2}$ \rf{eq:res}, which was purely classical, this nonlinear observable is essentially quantum. 

Now in contrast, when the motion is under the influence of random noise we solve the equation \rf{eq:schr4}. Its solution gives
\be \overline{ \VEV{x}^2} = \frac{2 \lambda  \Delta^3 m}{\sqrt{\pi} } \int_0^1 dx \frac{(1+x)^2}{ \left( \frac{2 \Delta^2
   m^2 \left(2 \Delta ^2+\ell ^2\right)}{t^2} +x^2\right)^{3/2}}.
   \ee
Again evaluating it for small $t \ll m \Delta^2$ gives 
\be \label{eq:noise} \overline{\VEV{x}^2}    \approx \frac{7  \lambda t^3}{3\sqrt{2 \pi}  m^2 \left(2 \Delta^2 + \ell^2 \right)^{3/2}}.
   \ee
Unlike the result \rf{eq:meas}, here the behavior is entirely classical and in fact has the same dependence on time as the linear result \rf{eq:res}. Note that quite interestingly, both 
the random measurement result \rf{eq:meas} and random noise result \rf{eq:noise} allow for the limit $\ell \rightarrow 0$. This cannot be said about the 
behavior of \rf{eq:res}. To make it entirely clear, we observe that for the choice of $g$ made here \rf{eq:g} $W''(0)=-1/(\sqrt{2\pi} \ell^3)$, therefore 
\be \VEV{x^2} = \Delta^2 + \frac{\hbar^2 t^2}{4 m^2 \Delta^2}  + \frac{ \lambda t^3}{3 \sqrt{2\pi} m^2 \ell^3} .
\ee
Clearly here the limit $\ell \rightarrow 0$ is not possible. This is related to the fact that measuring the particle on an infinitesimally small interval gives it undetermined velocity, in accordance with the
uncertainty principle. 

Finally we note that under random measurements the difference
\be \overline{\VEV{x^2}} - \overline{\VEV{x}^2} \approx \Delta^2 -\frac{2 \sqrt{2} \Delta^4 \lambda t}{\hbar^2 \sqrt{\pi} \left( 2 \Delta^2+ \ell^2 \right)^{3/2}},
\ee
where the approximate relation is given for a small time $t \ll m \Delta^2$ actually shrinks at first as a function of $t$ before growing again. The initial growth of $\overline{\VEV{x}^2}$ suppresses quantum spreading
for small enough $t$. This however does not happen for the motion with random noise. 

\acknowledgements
The author is grateful to L. Radzihovsky for many comments while this work was being completed, to I. Gornyi for discussing the definition and the importance of nonlinear observables and to N. Paul for bringing up the references \cite{Golubovic1991,Rosenbluth1992}. This work was supported by
 the Simons Collaboration on Ultra-Quantum Matter,
which is a grant from the Simons Foundation (651440).



\appendix

\section{Perturbative solution of the replicated Schr\"odinger equation}
\label{sec:a}
Let us now address the question of a randomly measured particle with the replica approach which allows to compute products of quantum mechanical averages, subsequently averaged over measurements.  We would like to solve the equations \rf{eq:schr3} and \rf{eq:schr4}. Here we present their perturbative solution for $\lambda$ small. 
\subsection{Higher order correlators: replicas}
We begin by writing down again the equation we would like to solve. Here and below we change notations from $x_{\alpha}^\sigma$ to $x_{\alpha, \sigma}$. 
\begin{eqnarray} \label{eq:eqw} i \frac{\d \rho}{d t} + \frac{1}{2m} \sum_{\alpha=1}^n \left( \frac{\d^2\rho}{\d x_{\alpha,+}^2} -  \frac{\d^2\rho}{\d x_{\alpha,-}^2} \right) = && \cr
 i \frac{\lambda}{2} \sum_{\alpha=1}^n \sum_{\beta=1}^n \sum_{\sigma,\sigma'=\pm}
W \left( x_{\alpha, \sigma} - x_{\beta, \sigma'} \right) \rho. &&
\end{eqnarray}
This needs to be solved with the initial condition, set at $t=0$,
\be \rho(0) = \frac{1}{\left( 2 \pi \Delta^2\right)^{n/2}  }  \exp \left\{ - \frac m 2 \sum_{\alpha=1}^n \left( \frac{x_{\alpha,+}^2}{2 \Delta^2 m } +  \frac{x_{\alpha,-}^2}{2 \Delta^2 m }   
\right)    \right\}. 
\ee
This choice of the function $W(x)$ simplifies the calculations,
\be W(x) = \frac{1}{\sqrt{2 \pi} \ell} e^{-\frac{x^2}{2 \ell^2}}.
\ee
If $\lambda=0$, then the following function solves this equation
\begin{eqnarray} \rho_0 = \frac{1}{\left( 2 \pi \Delta^2+ \frac{\pi t^2}{2 \Delta^2 m^2} \right)^{n/2}  } \times && \cr \exp \left\{ - \frac m 2 \sum_{\alpha=1}^n \left( \frac{x_{\alpha,+}^2}{2 \Delta^2 m + i t} +  \frac{x_{\alpha,-}^2}{2 \Delta^2 m - i t}   
\right)    \right\}. &&
\end{eqnarray}
We write $\rho\approx \rho_0 + \rho_1$, where $\rho_1$ satisfies
\begin{eqnarray} i \frac{\d \rho_1}{\d t} +\frac{1}{2m} \sum_{\alpha=1}^n \left( \frac{\d^2\rho_1}{\d x_{\alpha,+}^2} -  \frac{\d^2\rho_1}{\d x_{\alpha,-}^2} \right) = 
&& \cr
\frac{i \lambda}{2} \sum_{\alpha=1}^n \sum_{\beta=1}^n \sum_{\sigma,\sigma'=\pm}
W \left( x_{\alpha, \sigma} - x_{\beta, \sigma'} \right) \rho_0.&&
\end{eqnarray}
At $t=0$, $\rho_1=0$. 
To solve this we Fourier transform both sides. We arrive at
\be i \frac{\d \rho_1}{\d t} - \frac{1}{2m} \sum_{\alpha=1}^n \left( k_{\alpha,+}^2 - k_{\alpha_-}^2 \right) \rho_1 = \frac{i \lambda}{2} X(t).
\ee Here $X(t)$ will be determined below.
We solve this by writing
\be \rho_1 = C(t) \exp \left\{ {-i t \sum_{\alpha=1}^n \frac{k_{\alpha,+}^2-k_{\alpha,-}^2}{2m}} \right\}.
\ee
Here
\be C(t) = \frac{\lambda}{2} \int_0^t d\tau \, X(\tau) \exp \left\{ {i \tau \sum_{\alpha=1}^n \frac{k_{\alpha,+}^2-k_{\alpha,-}^2}{2m}} \right\},
\ee
thus
\begin{eqnarray} \label{eq:rho1} \rho_1 =\frac{ \lambda }{2}  \exp \left\{ {-i t \sum_{\alpha=1}^n \frac{k_{\alpha,+}^2-k_{\alpha,-}^2}{2m}} \right\} \times && \cr
\int_0^t d\tau \, X(\tau) \exp \left\{ {i \tau \sum_{\alpha=1}^n \frac{k_{\alpha,+}^2-k_{\alpha,-}^2}{2m}} \right\}. &&
\end{eqnarray}
Let us figure out $X$. It is given by the sum of the terms $X_{\alpha,\beta}^{\sigma,\sigma'}$ defined by
\begin{eqnarray}  X_{\alpha,\beta}^{\sigma,\sigma'} = \int dx_1 \dots dx_n \, \rho_0 \, W\left(x_{\alpha,\sigma}-x_{\beta,\sigma'} \right) \times && \cr
e^{-i \sum_{\alpha=1}^n \left( x_{\alpha,+} k _{\alpha,+} + x_{\alpha, -} k_{\alpha,-} \right) } .&&
\end{eqnarray}
Substituting and carrying out Gaussian integration, we find terms of three types. 
\begin{widetext}
\be X^{+-}_{\alpha,\beta}= \frac{\left( 8 \pi \Delta^2 \right)^{n/2}}{\sqrt{2 \pi} \sqrt{\ell^2+ 4 \Delta^2}} \exp \left\{-\frac{ \left( k_{\alpha,+} + k_{\beta,-} \right)^2 \left( \tau^2  + 4 m^2 \Delta^4  \right) +
2 m^2 \Delta^2 \ell^2 \left(k_{\alpha,+}^2+k_{\beta,-}^2\right)+
i m \ell^2 \tau \left(k_{\alpha,+}^2-k_{\beta,-}^2\right)  }{2 m^2 \left( 4 \Delta^2 +\ell^2 \right)}
\right\}  \times
\ee
$$ \prod_{\substack{ \gamma\not = \alpha \\ \delta \not = \beta} } \exp \left\{ - \frac{k_{\gamma,+}^2 \left( 2 m \Delta^2 + i \tau \right) + k_{\delta, -}^2 \left( 2 m \Delta^2 - i \tau \right)}{2 m} \right\},
$$
\be X^{++}_{\alpha, \beta} =\frac{\left(8 \pi \Delta^2 \right)^{n/2}}{ \sqrt{2 \pi} \sqrt{\ell^2+\frac{2 i \tau}{m}+4  \Delta^2  }}
\exp\left\{  -\frac{(k_{\alpha,+} + k_{\beta,+})^2 \left( 2 m \Delta^2 + i \tau \right)^2 + m \ell^2\left(2 m \Delta^2 + i \tau \right) (k_{\alpha,+}^2+k_{\beta,+}^2) }{2m (2 i\tau + m (4 \Delta^2 + \ell^2 ))}
\right\} \times 
\ee 
$$\prod_{\substack{ \gamma\not = \alpha \\ \gamma \not = \beta} } \exp \left\{ - \frac{k_{\gamma,+}^2 \left( 2 m \Delta^2 + i \tau \right) }{2 m} \right\}
\prod_\gamma \exp \left\{ - \frac{k_{\gamma,-}^2 \left( 2 m \Delta^2 - i \tau \right) }{2 m} \right\},
$$
\be X^{--}_{\alpha, \beta} =\frac{\left(8 \pi \Delta^2 \right)^{n/2}}{ \sqrt{2\pi} \sqrt{\ell^2-\frac{2 i \tau}{m}+4  \Delta^2  }}
\exp\left\{  -\frac{(k_{\alpha,-} + k_{\beta,-})^2 \left( 2 m \Delta^2 - i \tau \right)^2 + m \ell^2\left(2 m \Delta^2 - i \tau \right) (k_{\alpha,-}^2+k_{\beta,-}^2) }{2m (-2 i\tau + m (4 \Delta^2 + \ell^2 ))}
\right\} \times 
\ee 
$$\prod_{\substack{ \gamma\not = \alpha \\ \gamma \not = \beta} } \exp \left\{ - \frac{k_{\gamma,-}^2 \left( 2 m \Delta^2 - i \tau \right) }{2 m} \right\}
\prod_\gamma \exp \left\{ - \frac{k_{\gamma,+}^2 \left( 2 m \Delta^2 + i \tau \right) }{2 m} \right\}.
$$
Now for the purpose of calculating $\rho_1$, we need to conjugate $X$ with the appropriate factors, as in \rf{eq:rho1}. Defining
\be Y =   \exp \left\{ {-i t \sum_{\alpha=1}^n \frac{k_{\alpha,+}^2-k_{\alpha,-}^2}{2m}} \right\}X(\tau) \exp \left\{ {i \tau \sum_{\alpha=1}^n \frac{k_{\alpha,+}^2-k_{\alpha,-}^2}{2m}} \right\}
\ee
we find
\be Y^{+-}_{\alpha,\beta}  = \frac{\left( 8 \pi \Delta^2 \right)^{n/2}}{\sqrt{2 \pi} \sqrt{\ell^2+ 4 \Delta^2}} \exp \left\{-\frac{ \left( k_{\alpha,+} + k_{\beta,-} \right)^2 \left( \tau^2  + 4 m^2 \Delta^4  \right) +
2 m^2 \Delta^2 \ell^2 \left(k_{\alpha,+}^2+k_{\beta,-}^2\right)  }{2 m^2 \left( 4 \Delta^2 +\ell^2 \right)}
\right\}  \times
\ee
$$ \exp \left\{ \left( \frac{2 i \Delta^2 \tau}{4 m \Delta^2 + m \ell^2} - \frac{i t}{2m} \right) \left(k_{\alpha,+}^2- k_{\beta,-}^2 \right) \right\} \times \prod_{\substack{ \gamma\not = \alpha \\ \delta \not = \beta} } \exp \left\{ - \frac{k_{\gamma,+}^2 \left( 2 m \Delta^2 + i t \right) + k_{\delta, -}^2 \left( 2 m \Delta^2 - i t \right)}{2 m} \right\},
$$
\be Y^{++}_{\alpha, \beta} =\frac{\left(8 \pi \Delta^2 \right)^{n/2}}{ \sqrt{2 \pi} \sqrt{\ell^2+\frac{2 i \tau}{m}+4  \Delta^2  }}
\exp\left\{  -\frac{ \left(4 m^2 \Delta^4+\tau^2 + 2 m^2 \ell^2 \Delta^2 \right)(k_{\alpha,+}^2+k_{\beta,+}^2) + 2 k_{\alpha,+} k_{\beta,+} \left( 2 m \Delta^2 + i \tau \right)^2}{2m (2 i\tau + m (4 \Delta^2 + \ell^2 ))}
\right\} \times 
\ee 
$$\exp \left\{ -i t \frac{k_{\alpha,+}^2 + k_{\beta,+}^2}{2m} \right\} \prod_{\substack{ \gamma\not = \alpha \\ \gamma \not = \beta} } \exp \left\{ - \frac{k_{\gamma,+}^2 \left( 2 m \Delta^2 + i t \right) }{2 m} \right\}
\prod_\gamma \exp \left\{ - \frac{k_{\gamma,-}^2 \left( 2 m \Delta^2 - i t \right) }{2 m} \right\},
$$
\be Y^{--}_{\alpha, \beta} =\frac{\left(8 \pi \Delta^2 \right)^{n/2}}{ \sqrt{2 \pi} \sqrt{\ell^2-\frac{2 i \tau}{m}+4  \Delta^2  }}
\exp\left\{  -\frac{ \left(4 m^2 \Delta^4+\tau^2 + 2 m^2 \ell^2 \Delta^2 \right)(k_{\alpha,-}^2+k_{\beta,-}^2) + 2 k_{\alpha,-} k_{\beta,-} \left( 2 m \Delta^2 - i \tau \right)^2}{2m (-2 i\tau + m (4 \Delta^2 + \ell^2 ))}
\right\} \times 
\ee 
$$\exp \left\{ i t \frac{k_{\alpha,-}^2 + k_{\beta,-}^2}{2m} \right\} \prod_{\substack{ \gamma\not = \alpha \\ \gamma \not = \beta} } \exp \left\{ - \frac{k_{\gamma,-}^2 \left( 2 m \Delta^2 - i t \right) }{2 m} \right\}
\prod_\gamma \exp \left\{ - \frac{k_{\gamma,+}^2 \left( 2 m \Delta^2 + i t \right) }{2 m} \right\}.
$$
\end{widetext}
In terms of these, the correction to the density matrix reads
\begin{eqnarray} \rho_1 &=&
\lambda \left[ \int_0^t d\tau \, \sum_{\alpha < \beta} \left( Y^{+-}_{\alpha \beta} + Y^{+-}_{\beta \alpha}  +Y^{++}_{\alpha \beta} + Y^{--}_{\alpha \beta} \right)+ \right.  \cr && \left. \sum_{\alpha} Y^{+-}_{\alpha \alpha} \right]. 
\end{eqnarray}

Now we introduce classical and quantum coordinates,  with symmetric normalization
\be x_{cl} = \frac{x_++x_-}{\sqrt{2}}, \ x_q = \frac{x_+-x_-}{\sqrt{2}}, 
\ee
\be k_{cl} = \frac{k_++k_-}{\sqrt{2}}, \ k_q = \frac{k_+-k_-}{\sqrt{2}} 
\ee
Importantly,
\be x_+ k_+ + x_- k_- = x_{cl} k_{cl}+ x_q k_q, 
\ee
\be k_+ = \frac{k_{cl}+k_q}{\sqrt{2}}, \ k_- = \frac{k_{cl}- k_q }{\sqrt{2}}.
\ee
\subsection{Nomalizing the density matrix}
We would now like to normalize the density matrix. To do that, we set $k_{cl}=0$ and integrate over $k_q/(2 \pi)$. This is equivalent to setting $k_{\alpha,+}=-k_{\alpha,-}=k_\alpha/\sqrt{2}$ and integrating over $k_\alpha/(2\pi)$. After that we still need to normalize by dividing by $2^{n/2}$. 

Let us see that in more detail. Normalizing the density matrix means computing
\be {\cal N} = \int dx \, \rho(x,x).
\ee
The norm of the density matrix is supposed to be $1$, but the solutions of the equation \rf{eq:eqw} is not automatically normalized, so the density matrix inferred from that equation must be normalized by dividing by ${\cal N}$. Technically this is because we have not normalized the probability distribution for $V$ and ignored factors such as $W(0)$ (self interactions among the replicas) which violates the conservation of the norm of $\rho$. Diving by ${\cal N}$ fixes this issue.  
We have defined
\begin{eqnarray} \rho(k_+, k_-) = \int dx_- dx_+ \, \rho(x_+, x_-) e^{-i k_- x_- - i k_+ x_+} =&& \cr
 \int dx_{cl} dx_q \, \rho((x_{cl}+x_q)/\sqrt{2}, (x_{cl}- x_q)/\sqrt{2}) e^{-i k_{cl} x_{cl}- i k_q x_q}. && \cr &&
\end{eqnarray}
Setting $k_{cl}=0$ and integrating over $k_q$ gives
\begin{eqnarray} \int \frac{dx_{cl} dx_q dk_{q}}{2\pi} \, \rho((x_{cl}+x_q)/\sqrt{2}, (x_{cl}- x_q)/\sqrt{2}) \, e^{-i k_q x_q} = && \cr
 \int dx_+ dx_- \, \rho(x_+,x_-) \,\sqrt{2} \delta(x_+-x_-) = \sqrt{2} {\cal N}.  && \cr && 
\end{eqnarray}

Applied to the coefficients $Y$ above, 
with the definition
\be Z = \frac{1}{2^{n/2}} \lim_{k_{cl} \rightarrow 0} \int \prod_{\alpha=1}^n \frac{d k_{\alpha,q}}{(2\pi)^n} Y,
\ee
this procedure gives
\be Z^{+-}_{\alpha, \beta} = \frac{m \Delta}{\sqrt{\pi}} \frac{1}{\sqrt{2 m^2 \Delta^2 ( 2 \Delta^2 + \ell^2) + \tau^2}}, \ \alpha < \beta,
\ee
\be Z^{+-}_{\alpha, \alpha} =  \frac{1}{\sqrt{2 \pi} \ell},
\ee
\be Z^{++}_{\alpha,\beta} = Z^{--}_{\alpha, \beta} = \frac{m \Delta}{\sqrt{\pi}} \frac{1}{\sqrt{2 m^2 \Delta^2 ( 2 \Delta^2 + \ell^2) + \tau^2}}.
\ee
Thus we find
\begin{eqnarray}  {\rm tr} \, \rho_1 & = & 2 n(n-1) \lambda \int_0^t d\tau \frac{m \Delta}{\sqrt{\pi}} \frac{1}{\sqrt{2 m^2 \Delta^2 ( 2 \Delta^2 + \ell^2) + \tau^2}} \cr
&& + \frac{n \lambda t}{\sqrt{2\pi} \ell}.
\end{eqnarray}
\subsection{Computing the relevant correlator}
We would now like to compute the correlator. To do that, we differentiate over $k_{cl,1}$ and $k_{cl,2}$, and then set all $k_{cl}$ to zero. After that we integrate over each $k_q/(2 \pi)$ and multiply by $-1/(2 \cdot 2^{n/2})$. Here is a check that this is the right procedure:
\be \int \frac{dk_q}{(2\pi)}  \left. \frac{\partial }{\partial k_{cl}} \rho \right|_{k_{cl}=0} = 
\ee
$$ \frac{\partial }{\partial k_{cl}}   \int \frac{ dx_+ dx_- dk_q}{2\pi}  \rho(x_+, x_-) e^{-i k_{cl} \frac{x_+ + x_-}{\sqrt{2}} - i k_q \frac{x_+-x_-}{\sqrt{2}} } = $$
$$ - \int dx_+ dx_- \rho(x_+, x_-) \sqrt{2} \delta(x_+ - x_-) \frac{x_+ + x_- }{\sqrt{2}} = 
$$
$$- \int dx \rho(x,x) \sqrt{2} \cdot \sqrt{2} x.
$$
We see a factor of $\sqrt{2}$ per correlator (hence, division by $2$) and a factor of $\sqrt{2}$ per a pair of degrees of freedom, hence division by $2^{n/2}$. 

We see that the only nonzero contribution comes from the terms where $\alpha=1$, $\beta=2$. If this is not so, suppose $\beta=3$. Then the derivative over $k_{3,cl}$ followed by setting it to zero brings down $k_{3,q}$. Integrating over $k_{3,q}$ is then zero as the exponent is symmetric under the change of sign of $k_{3,q}$. 

The exception is the $\alpha=1$, $\beta=2$ terms. To compute $S^{+-}_{12}$ use the following formula
$$ \frac 1 Z \int \frac{dk_{1,q} d k_{2,q}}{(2\pi)^2}  \frac{\d^2} {\d k_{1,cl} \d k_{2,cl}} \exp\left(-\frac 1 2 \left( a k_{1,+}^2 + \right. \right.
$$
$$ \left. \left.   b k_{2,-}^2+2 c k_{1,+} k_{2,-} + d k_{1,-}^2+f k_{2,+}^2 \right) \bigg) \right|_{k_{1,cl}, k_{2,cl}=0}
= $$
\be \label{eq:ans1} \frac{2 c d f}{c^2-(a+d)(b+f)},
\ee
where
$$ Z =  \int \frac{dk_{1,q} d k_{2,q}}{(2\pi)^2} \exp\left(-\frac 1 2 \left( a k_{1,+}^2 +  \right. \right. 
$$ $$ \left.  b k_{2,-}^2+2 c k_{1,+} k_{2,-} +d k_{1,-}^2+f k_{2,+}^2 \right) \bigg).
$$
 We find
 \begin{eqnarray} S^{+-}_{1,2}& =& \frac 1 2 Z^{+-}_{1,2} \frac{\left( t^2+4 m^2 \Delta^4\right) \left( \tau^2+4 m^2 \Delta^4 \right)}{4 m^2 \Delta^2 \left( 2 m^2 \Delta^2 (2 \Delta^2+\ell^2) + \tau^2 \right)}= \cr
&& \frac{\left( t^2+4 m^2 \Delta^4\right) \left( \tau^2+4 m^2 \Delta^4 \right)}{8\sqrt{\pi}  m \Delta \left( 2 m^2 \Delta^2 (2 \Delta^2+\ell^2) + \tau^2 \right)^{3/2}}. 
 \end{eqnarray}
 Now to compute $S^{++}_{1,2}$ we use 
$$  \frac 1 Z \int \frac{dk_{1,q} d k_{2,q}}{(2\pi)^2}  \frac{\d^2} {\d k_{1,cl} \d k_{2,cl}} \exp\left(-\frac 1 2 \left( a k_{1,+}^2 +  \right. \right.
$$
$$ \left. \left.  b k_{2,+}^2+2 c k_{1,+} k_{2,+} + d k_{1,-}^2+f k_{2,-} ^2\right) \bigg) \right|_{k_{1,cl}, k_{2,cl}=0}
= $$
$$ \frac{2 c d f}{c^2-(a+d)(b+f)},
$$
where
$$ Z =  \int \frac{dk_{1,q} d k_{2,q}}{(2\pi)^2} \exp\left(-\frac 1 2 \left( a k_{1,+}^2 +  \right. \right. 
$$
$$ \left. b k_{2,+}^2+2 c k_{1,+} k_{2,+} + d k_{1,-}^2+f k_{2,-}^2 \right) \bigg).
$$
That is, this is explicitly the same expression as \rf{eq:ans1}. 
 
 This gives
$$  S^{++}_{1,2}= \frac 1 2 Z^{++}_{1,2} \frac{\left( it+2  m \Delta^2 \right)^2 \left( 2m \Delta^2 + i \tau\right)^2}{4 m^2 \Delta^2 \left( 2 m^2 \Delta^2 \left( 2 \Delta^2+\ell^2 \right) + \tau^2 \right)} =
$$
$$
 \frac{\left( it+2  m \Delta^2 \right)^2 \left( 2m \Delta^2 + i \tau\right)^2}{8\sqrt{\pi} m \Delta \left( 2 m^2 \Delta^2 \left( 2 \Delta^2+\ell^2 \right) + \tau^2 \right)^{3/2}} 
 $$ 
 This gives the answer for the quantity we are trying to compute
$$  \overline{ \VEV{x}^2} =   \lambda \int_0^t d\tau \left( 2S^{+-}+S^{++}+{S^*}^{++} \right) = $$
$$\frac{ \lambda }{2} \int_0^t d\tau \frac{ \left(t \tau -4 \Delta ^4 m^2\right)^2}{\sqrt{\pi } \Delta  m \left(2 \Delta ^2
   m^2 \left(2 \Delta ^2+\ell ^2\right)+\tau ^2\right)^{3/2}}.
  $$

 This integral is not very instructive. Let us rescale $\tau = t x$ to find
 \begin{eqnarray} \overline{ \VEV{x}^2} & =&  \frac{ \lambda t^2}{2 \sqrt{\pi } \Delta  m} \int_0^1 dx \frac{ \left(x -\frac{4 \Delta ^4 m^2}{t^2} \right)^2}{ \left(\frac{2 \Delta^2
   m^2 \left(2 \Delta ^2+\ell ^2\right)}{t^2} +x^2\right)^{3/2}} \cr & \approx & \frac{\lambda t^2}{2 \sqrt{\pi} \Delta m} \ln \left( \frac {t\sqrt{2} } {\sqrt{\Delta^2
   m^2 \left(2 \Delta ^2+\ell ^2\right)} } \right).
 \end{eqnarray}
 The last approximate equality works for large $t$.  Units: given that $\lambda$ has units of $\left[ \lambda \right] = {\rm kg}^2 \cdot {\rm m}^5 \cdot {\rm s}^{-3}$, restoring the Planck constant gives
 \be \overline{ \VEV{x}^2} = \frac{\lambda t^2}{2 \sqrt{\pi} \Delta m \hbar} \ln \left( \frac {\hbar t\sqrt{2} } {\sqrt{\Delta^2
   m^2 \left(2 \Delta ^2+\ell ^2\right)} } \right).
   \ee
 
 Let us now evaluate this at small $t$. We find
 \be \overline{ \VEV{x}^2 } \approx \frac{2 \sqrt{2} \Delta^4 \lambda t}{\hbar^2 \sqrt{\pi} \left( 2 \Delta^2+ \ell^2 \right)^{3/2}}.
 \ee

 \subsection{Now the case of random noise}
 
 If instead of random measurements we simply have a random potential, then we need to compute
$$   \lambda \int_0^t d\tau \left( 2S^{+-}-S^{++}-{S^*}^{++} \right) =  $$
$$\lambda \int_0^t d \tau \frac{4 \Delta ^3 m (t+\tau )^2}{2 \sqrt{\pi } \left(2 \Delta ^2 m^2 \left(2 \Delta
   ^2+\ell ^2\right)+\tau ^2\right)^{3/2}}.
  $$
       Changing variables in the same way $\tau = t x$, we find
       \begin{eqnarray} \overline{ \VEV{x}^2} & = & \frac{2 \lambda  \Delta^3 m}{\sqrt{\pi} } \int_0^1 dx \frac{(1+x)^2}{ \left( \frac{2 \Delta^2
   m^2 \left(2 \Delta ^2+\ell ^2\right)}{t^2} +x^2\right)^{3/2}}  \cr & \approx & \frac{ \lambda  \Delta }{\hbar \sqrt{\pi} m }  \frac{t^2}{  2 \Delta
   ^2+\ell ^2 }.
   \end{eqnarray}
   The last approximate equality is valid at large $t$, and $\hbar$ in it is again inserted as appropriate. 
   
   At small $t$ we find
   \be \overline{ \VEV{x}^2} \approx \frac{7  \lambda t^3}{3\sqrt{2 \pi}  m^2 \left(2 \Delta^2 + \ell^2 \right)^{3/2}}.
   \ee
   Here there is no $\hbar$, so the result is classical.

 \subsection{Computing the linear correlator}
 
For completeness, let's compute the linear correlator $\VEV{x^2}$. The only term that contributes is  $Y^{+-}_{11}$. We differentiate it over $k_{cl,1}$ twice, set $k_{cl,1}$ to zero, and integrate over $k_{q,1}/(2\pi)$, multiplying by a factor of $-1/(2\cdot2^{n/2})$. Other terms will go to zero as $n \rightarrow 1$ for combinatorial reasons, so there are not needed. 
 
Carrying out this procedure gives
\be  \int_0^t d \tau \,  \lambda \frac{t^2 \left(4
   \Delta ^2+\ell ^2\right) +4 \Delta ^2 \left(\Delta ^2 m^2 \ell ^2+\tau ^2\right)-8 \Delta ^2 \tau  t}{4 \sqrt{2\pi}  \Delta ^2 m^2
   \ell ^3}.
   \ee
Changing variables as always gives
$$ \frac{\lambda t^3}{4 \sqrt{2\pi} \Delta^2 m^2 \ell^3} \int_0^1 dx \, \left( 4
   \Delta ^2+\ell ^2 + 4 \Delta ^2 \left(\frac{\Delta ^2 m^2 \ell ^2}{t^2} +x^2 \right) \right.$$
   $$ -8 \Delta ^2 x \bigg) = \frac{ \Delta^2 \lambda t}{\sqrt{2 \pi} \ell} + \frac{\lambda t^3 \left(4 \Delta^2+3 \ell^2 \right)}{12 m^2 \sqrt{2 \pi} \Delta^2 \ell^3}.
   $$
Finally, the correlator is given by
\begin{eqnarray} \VEV{x^2} & =& \left( \Delta^2 + \frac{t^2}{4 m^2 \Delta^2} \right) \left(1- \frac{ \lambda t}{\sqrt{2\pi} \ell} \right)+\frac{ \Delta^2 \lambda t}{\sqrt{2 \pi} \ell} +
\cr && \frac{\lambda t^3 \left(4 \Delta^2+3 \ell^2 \right)}{12 m^2 \sqrt{2 \pi} \Delta^2 \ell^3} = \frac{\lambda  t^3}{3 \sqrt{2 \pi} m^2 \ell^3}.
\end{eqnarray}
This matches \rf{eq:res}.


\bibliography{library.bib}

\end{document}